\documentclass{article}
\usepackage{array}
\usepackage{color}
\usepackage{graphicx}
\usepackage{authblk}

\title{Dimensional reduction and adaptation-development-evolution relation in evolved biological systems}

\author{Kunihiko Kaneko}
\date{}

\affil{Niels Bohr Institute, University of Copenhagen, Blegdamsvej 17, 2100 Copenhagen, Denmark
}

\begin{document}
\maketitle
\begin{abstract}
Living systems are complex and hierarchical, with diverse components at different scales, yet they sustain themselves, grow, and evolve over time. How can a theory of such complex biological states be developed? Here we note that for a hierarchical biological system to be robust, it must achieve consistency between micro-scale (e.g. molecular) and macro-scale (e.g. cellular) phenomena. This allows for a universal theory of adaptive change in cells based on biological robustness and consistency between cellular growth and molecular replication.
Here, we show how adaptive changes in high-dimensional phenotypes (biological states) are constrained to low-dimensional space, leading to the derivation of a macroscopic law for cellular states. The theory is then extended to evolution, leading to proportionality between evolutionary and environmental responses, as well as proportionality between phenotypic variances due to noise and due to genetic changes. The universality of the results across several models and experiments is demonstrated. Then, by further extending the theory of evolutionary dimensional reduction to multicellular systems, the relationship between multicellular development and evolution, in particular, the developmental hourglass, is demonstrated. Finally, the possibility of collapse of dimensional reduction under nutrient limitation is discussed.
\end{abstract}

\section{Introduction}

Biological systems generally consist of a huge number of components, and are hierarchical: cells consist of diverse molecules, organisms consist of diverse cells, and ecosystems consist of diverse species\cite{KK-book}. In order to have robustness in a biological system, there must be some consistency across levels; for example, in order for a cell to grow, almost all the components in it must be produced in proportion to their abundances; otherwise the cellular state is not robust. 
In other words, the system has to satisfy {\sl macro-micro consistency}; the cell growth at a macroscopic level need to keep consistency with the replication rate of each of constituent molecules at a microscopic level, when a cell achieves a state with robust growth. Likewise, for a multicellular organism, multicellular growth at an organism level needs to keep some consistency with dynamics of each cell. 
In the field of {\sl universal biology} that we intend to establish, we seek to uncover universal laws in adaptation, development, and evolution by noting robustness as well as plasticity to adapt to environmental changes.  Here, we briefly review recent advances in the field, focusing on the evolutionary dimensional reduction of high-dimensional biological states, as well as the correlation between the changes in evolution and adaptation or development. Last, possible violation of the dimension reduction is discussed in relationship with the transition to dormant (non-growing) state of microorganisms.

\section{Dimensional reduction as a consequence of robustness}

\begin{figure}
\begin{center}
\includegraphics[width=10cm]{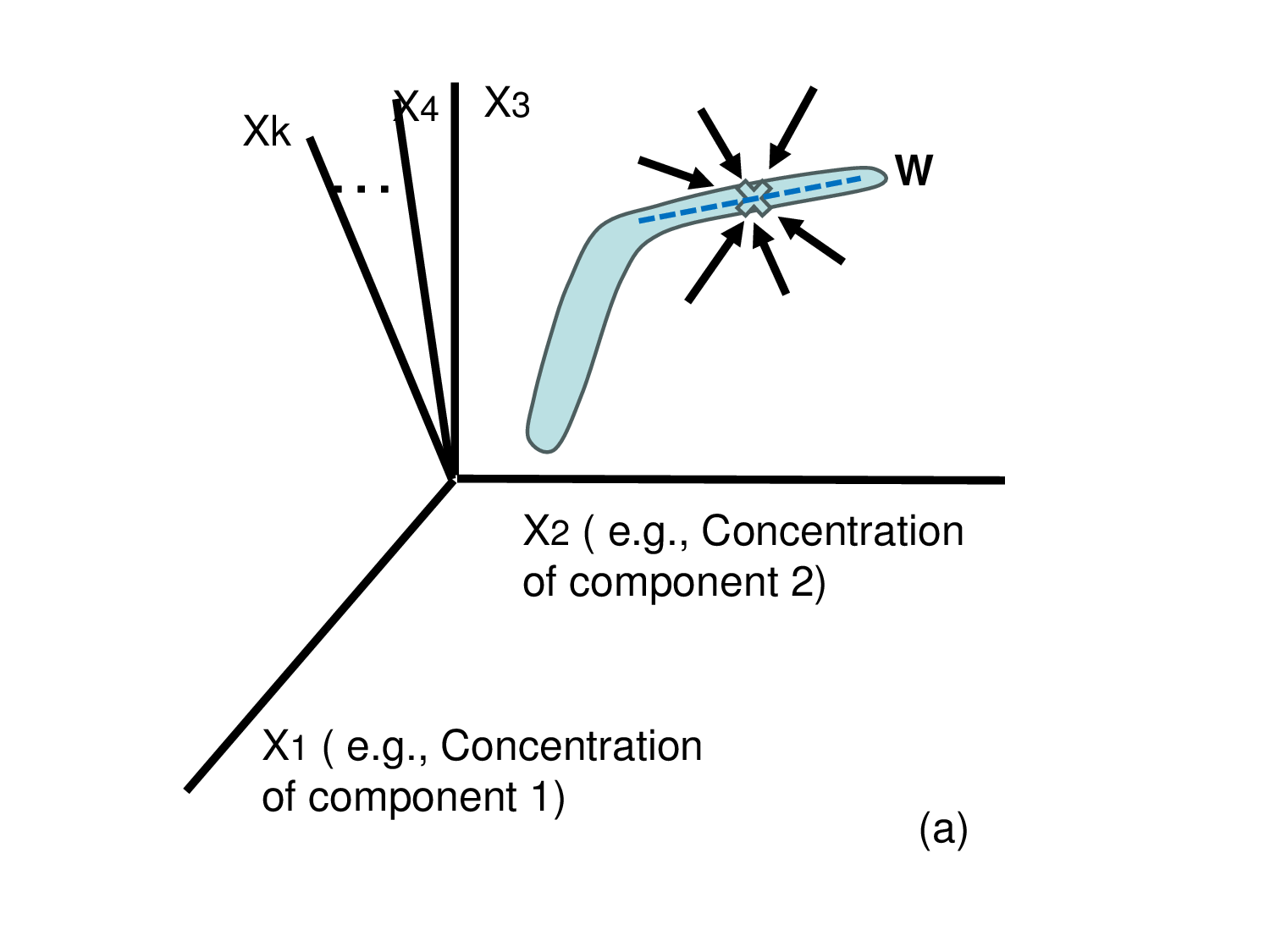}
\includegraphics[width=10cm]{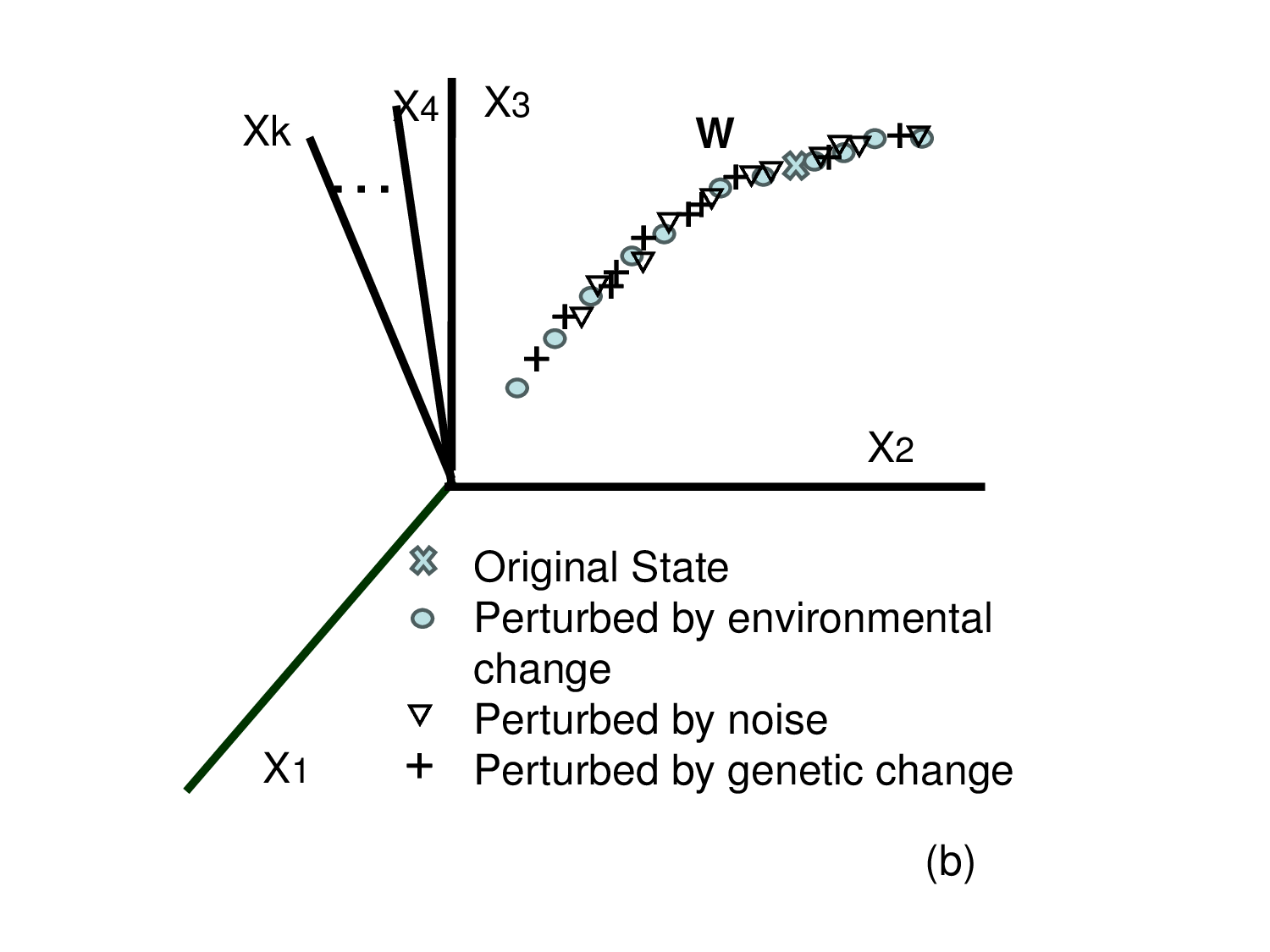}
\end{center}
\caption{
Schematic representation of low-dimensional constraint of phenotype in the state space. (a)Robustness of phenotype to perturbation and plasticity along the environmental condition leads to dimension reduction against different environmental conditions. (b) The responses of phenotypes to environmental, genetic (evolutionary) changes and noise are constrained in the same low-dimensional manifold (space).
}
\end{figure}

Recalling the diversity of components in a biological system, its state is high-dimensional: In a cell, there are typically thousands of components.  Then, the cellular state to be described by the concentration of each component requires a quite high-dimensional space; for a bacterial cell, about a few thousand protein or mRNA species (besides more metabolites), and the dimension of the state space exceeds at least a few thousand. Recent experimental and simulation studies, however, suggest that a drastic reduction in the dimensionality of the state change (phenotypes) in response to environmental changes has taken place[2-7], 
whereas the dimensional reduction of evolved phenotypes has been observed extensively in a variety of contents, as will be discussed in \S 6.

 First, it is noted that the steady state imposes a constraint on the phenotypic changes so that the cell growth rate and replication rates of molecules are balanced\cite{Mu}, but it alone cannot explain the experimentally observed dimensional reduction in the changes of many components across different environmental conditions. 

Then, we simulated a cell model consisting of a large number of components with catalytic-reaction network \cite{CFKK-PRE}, in which phenotypes are generated as a result of intracellular reaction dynamics, where the abundances of each component change with catalytic reactions by taking up nutrient chemicals from the environment, leading to cellular growth.  
When we introduced arbitrarily chosen reaction dynamics into the model, no such reduction was observed, even when the steady growth condition of the cells was satisfied. In contrast, when these model cells were evolved to achieve higher cell growth under a given environmental condition, it is found that the phenotypic changes of the cells upon a variety of environmental perturbations are constrained within a low-dimensional space.

The origin of such dimensional reduction has been explained as follows\cite{CFKK-PRE,KKCF-Rev}:
Consider high-dimensional dynamical systems that shape an adapted phenotype.  In general,  such dynamics are subject to various perturbations.  In fact, intracellular reactions are noisy because the number of molecules inside is not so large, and the environmental conditions also fluctuate. Then the fitted phenotype state should be robust to such perturbations. 

Accordingly, in the dynamics for the phenotype, there should be a strong attraction to the fitted state (attractor) from most directions in the high-dimensional state space. In terms of dynamical systems, this implies that there is strong attraction from most directions to this fitted attractor. However, if this strong attraction occurred from all directions, evolution to increase fitness would be difficult because the state would not be easily changed by genetic mutations that slightly alter the dynamical system (the flow in the state space). Then, along the direction of evolution, the phenotypes should be feasible to change. Accordingly, dominant changes in phenotypes (e.g., concentration of proteins) will be constrained along a one-dimensional curve (or within a few-dimensional space) in which evolution has progressed and will progress (see Fig. 1). 

In fact, we confirmed the above behavior, by the evolution simulation of a cell model with catalytic reaction dynamics mentioned above. By linearizing the dynamics around the steady state, one (or few) eigenvalues for the relaxation matrix (Jacobi matrix) is close to zero, as distinct from other eigenvalues which are much more negative. The evolutionary change has progressed along the eigenvector for the eigenvalue close to zero\cite{CFKK-PRE,Sato-KK1}. 

In addition to the cell model with catalytic reaction network, the dimensional reduction has also been confirmed in the evolution of gene-regulatory network models with mutual activation or suppression of gene expression levels\cite{Sato-KK2023}. In the gene-expression model, fitness is determined by the expression levels of given target genes, depending on environmental conditions.  In these models also, the dynamics involve many variables that provide the high-dimensional phenotypic dynamics, from which the fitness for evolution is determined.  In the numerical studies, it was found that after evolution, changes in most variables induced by a variety of different environmental perturbations are highly constrained along a one-dimensional manifold.

Here, with the dimensional reduction, the relaxation dynamics towards the steady state involves one (or few) slow modes. The evolution of such a slow mode is reasonable by the following argument: Consider a complex system with many interacting elements that have similar time scales. It will be more difficult to evolve such a system because the random mutations to each element (gene) would interfere with each other and cancel each other out.  Then, directed phenotypic changes by genetic changes would be more difficult, as the saying goes {\sl too many cooks spoil the broth}. On the other hand, if a few slow modes are segregated, they will work to control others. Then mutations in the genes for the slow modes will strongly affect fitness. In other words, the separation of a few slow modes, which is a result of evolution, also facilitates phenotypic evolution.

\section{Relationship between environmental and evolutionary responses}

Since the one-dimensional curve for the phenotypic changes in response to environmental changes follows the curve along which the evolution has progressed,  the responses due to environmental and evolutionary (genetic) changes will occur along the same low-dimensional manifold (see Fig. 1b). Thus, phenotypic changes caused by mutation are highly correlated with those caused by environmental or stochastic perturbations. This correlation was confirmed by the simulation of the cell models.

Here let us denote the change in phenotype $X_i$ (e.g., the log concentration of each component) due to environmental responses as $\delta X_i(E)$, and that due to genetic evolution as $\delta X_i(G)$. Since all phenotype changes are then constrained along the common one-dimensional curve, $\delta X_i(G) = \delta W(G) cos \theta _i$ and $\delta X_i(E) =\delta W(E) cos \theta_i$ will follow, where $\delta W$ is the change in the principal mode $W$ and $\theta _i$ is the angle between $W$ and $X_i$. Then, $\delta X_i(G)/\delta X_i(E) =\delta W(G)/\delta W(E)$ will be independent of each component $i$. ($cos\theta_i$ is the case in which the principal axis is linear. If it is curved, one can use a function $f(\theta_i)$ that represents the projection of $X_i$ with the principal axis, and the same result is obtained).

Now consider genetic evolution after applying an environmental stress that decreases the growth rate $\mu$ with $\delta \mu(E)<0$. Evolution restores the growth rate so that the change from the original unstressed state, $|\delta \mu (G)|$, should be less than $|\delta \mu(E)|$.  Here note that the growth rate $\mu$ of the cell is also a phenotype, and can be regarded as one of $X_i$'s. Hence, the relationship
\begin{equation}
\delta X_i(G)/\delta X_i(E) =\delta \mu (G)/\delta \mu (E)
\end{equation}
is expected to hold.

Since $|\delta \mu(G)|< |\delta \mu(E)|$, this equation implies that the stress-induced change in phenotype $\delta X_i(E)$ (e.g., the logarithm of the concentration of a protein or mRNA species) is reduced as evolution proceeds under that environmental stress, since $\delta X_i(G)/\delta X_i(E) <1$.  Thus, the long-term evolutionary response is to counteract the environmentally induced change of $X_i$. This {\sl homeostasis} by evolutionary adaptation is akin to the Le Chatelier principle in thermodynamics\cite{CFKK-Interface,KKCF-Rev}. In fact, eq. (1), together with the Le Chatelier-type responses, has been confirmed both in the simulation of cell models and in the evolution experiment of bacteria\cite{CFKK-Interface,Horinouchi2}. This Le Chatelier-type response can be understood by assuming that the fitness (growth rate) is described as a smooth, potential function of reciprocal variables representing environmental and genetic changes\cite{Sato-KK2023,KK-unpub}.

\section{Proportionality between the phenotypic variances by noise and by mutation}

As mentioned above, intracellular dynamics (or any biological process that shapes the phenotype) are rather noisy, and phenotypic fluctuations are often rather large in cells\cite{Elowitz,log-normal,Hashimoto}. Here, it should be recalled that in statistical physics, response and fluctuations are two sides of the same coin, as demonstrated by the fluctuation-response relationship\cite{Einstein,Kubo}.  In fact, we have previously shown an evolutionary fluctuation-response relationship in which the evolutionary response (speed) is proportional to the variance of the phenotype over isogenic individuals (clones) induced by noise, denoted as $V_{ip}$ (isogenic phenotypic variances)\cite{Sato}. 

In considering the phenotypic variance, however, there also exists another source that causes the phenotypic fluctuations. It is the change in genes that govern the gene expression dynamics(see Fig.2). In fact, in the standard population genetics, the genetic variance $V_g$, which is the variance of the phenotype due to the genetic distribution, is known to be proportional to the rate of evolution\cite{Fisher,Falconer}. Accordingly, the proportionality between the phenotypic variances due to noise $V_{ip}$ and due to genetic change $V_g$ is proposed, which is confirmed in the numerical simulation of the cell model over the course of evolution\cite{CFKK-JTB,KK-PLoS}.  

Now, recall the proportionality of phenotypic responses by external environmental changes and by genetic variation across all components. We can then expect the proportionality of two phenotypic variances of $X_i$ by noise and by mutation, across all components $i$. This can also be shown as follows: Let us denote the variance of the former by $V_{ip}(i)$ and that of the latter by $V_g(i)$. As shown in Fig. 1b, changes due to noise and genetic variation are constraiend along the common principal axis. 
Thus, phenotypic variances due to the former are given by $V_{ip} (i)\equiv \langle \delta X_i(E)^2 \rangle_{noise}$ and those due to the latter by $V_g(i)\equiv \langle (\delta X_i(G)^2 \rangle_{mutation}$,
where $\langle \cdots \rangle_{noise}$ and $\langle \cdots \rangle_{mutation}$ are the average over the distribution of phenotypes induced by noise and mutations, respectively.  
Recall that the relaxation of phenotypic changes is much slower along the principal axis $W$ in Fig.1, phenotypic fluctuations due to noise are nearly confined to this axis. Thus, the variance of each component $X_i$ is given by the mean of the variance in the variable $W$ (along the principal axis) as $V_{ip}(i) = \langle \delta W^2 \rangle_{noise} cos^2 \theta_i$ (due to noise) and $V_{g}(i) = \langle \delta W^2 \rangle_{mutation} cos^2 \theta_i$ (due to mutation). Hence,
\begin{math}
V_{ip}(i)/ V_{g}(i) = \langle \delta W^2 \rangle_{noise}/
\langle \delta W^2 \rangle_{mutation}
\end{math} follows. Accordingly
\begin{equation}
V_{ip}(i) \propto V_{g}(i) 
\end{equation} will hold over the components $i$.

This proportionality $V_{ip}(i)\propto V_g(i)$ over most of the components $i$ was confirmed by using the aforementioned cell model with catalytic reaction network \cite{CFKK-Interface}, as well as in the evolution of the gene regulatory network model\cite{KK-ESB,KK-JSP}, after the evolution had proceeded to achieve robustness.

\begin{figure}
\begin{center}
\includegraphics[width=12cm]{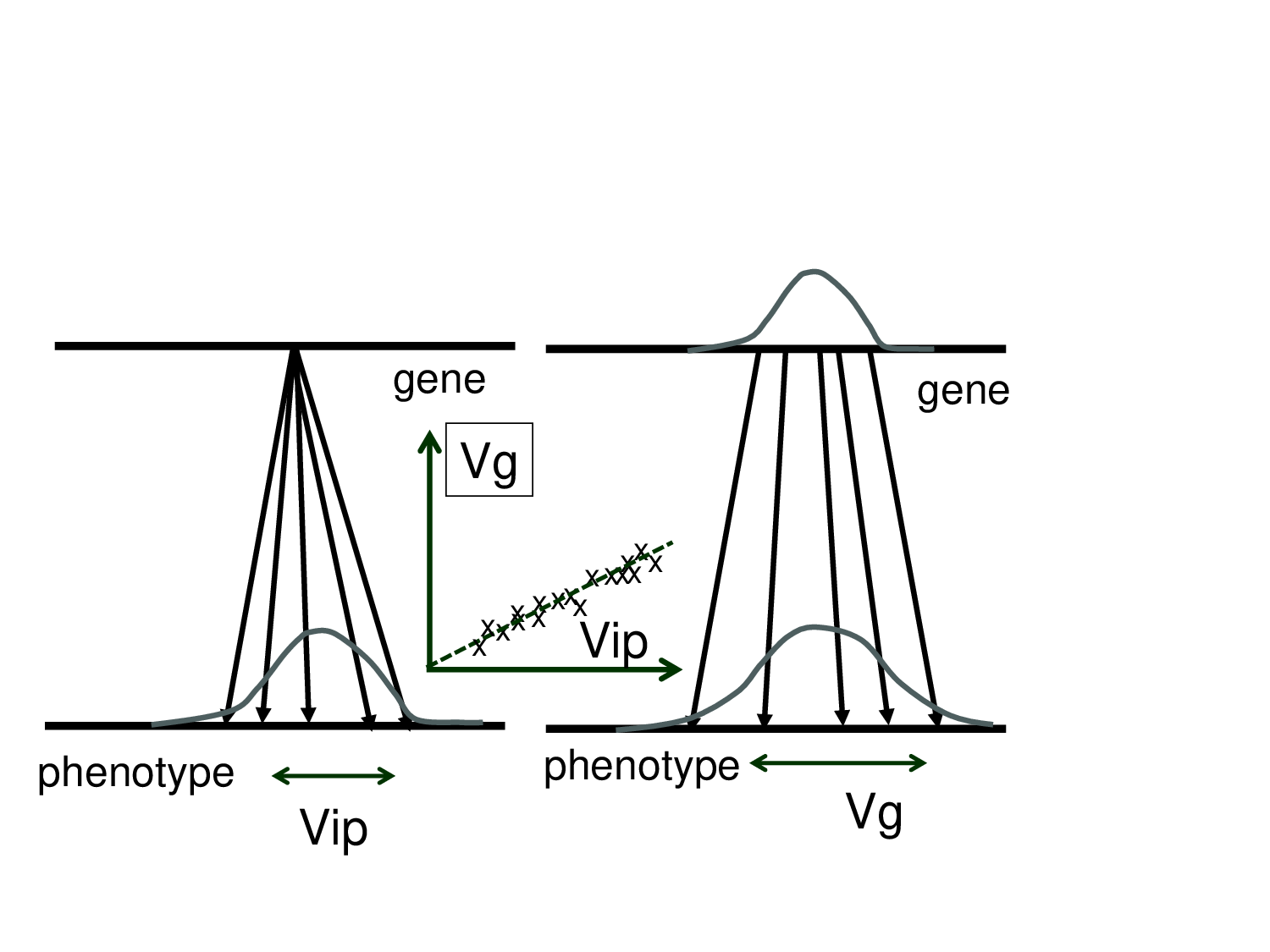}
\end{center}
\caption{
Schematic representation of  $V_{ip}(i)$ and $V_g(i)$. Both are the variance of phenotypes but the former is a result of noise across isogenic individuals, whereas the latter is due to the genetic distribution. Despite the difference in the origin of the variation, the two are highly correlated.
}
\end{figure}

Note that this result introduces the possibility of predicting which component $i$ is likely to evolve: changes due to noise are not inherited, but genetic variation is transmitted to offspring, leading to the evolutionary change. Accordingly, the resulting phenotypic variance $V_g(i)$ is proportional to the rate of evolution of that component, known as Fisher's {\sl fundamental theorem of natural selection} \cite{Fisher,Falconer}. Now that we have shown that $V_{ip}(i)$ is proportional to $V_g(i)$, the rate of evolution of $X_i$ is proportional to $V_{ip}(i)$. In other words, the greater the variance of component (trait) $i$ due to noise or environmental change (before genetic change), the more likely component (trait) $i$ is to evolve. In other words, the direction in which it is likely to evolve in phenotypic space can be predicted in advance of genetic evolution\cite{KKCF-Rev}. 

\section{Summary of evolutionary dimensional reduction and $V_g$-$V_{ip}$ law}

In summary, we have demonstrated the evolutionary dimensional reduction and 2-by-2 global proportionality of phenotypic changes between responses and fluctuations, and between perturbations due to environmental changes or  noise and genetic changes, as schematically shown in Fig. 3.

We expect that this evolutionary dimensional reduction and the resulting constraint on phenotypic evolution will be universal in evolved biological systems. In the theoretical argument and cell models, the prerequisites for the emergence of dimensional reduction are only the following two: (i) phenotypes with higher fitness are shaped by complex high-dimensional dynamical systems, which involve stochasticity and (ii) the fraction of such fit states is rare in the state space and in the genetic-rule space. These two features are necessary for our theoretical argument, while most biological systems satisfy them. 
Intuitively, a high fitness state satisfying macro-micro consistency must be rare, and once it is achieved, its conservation is favored, implying the requirement of robustness.

\begin{figure}
\begin{center}
\includegraphics[width=10cm]{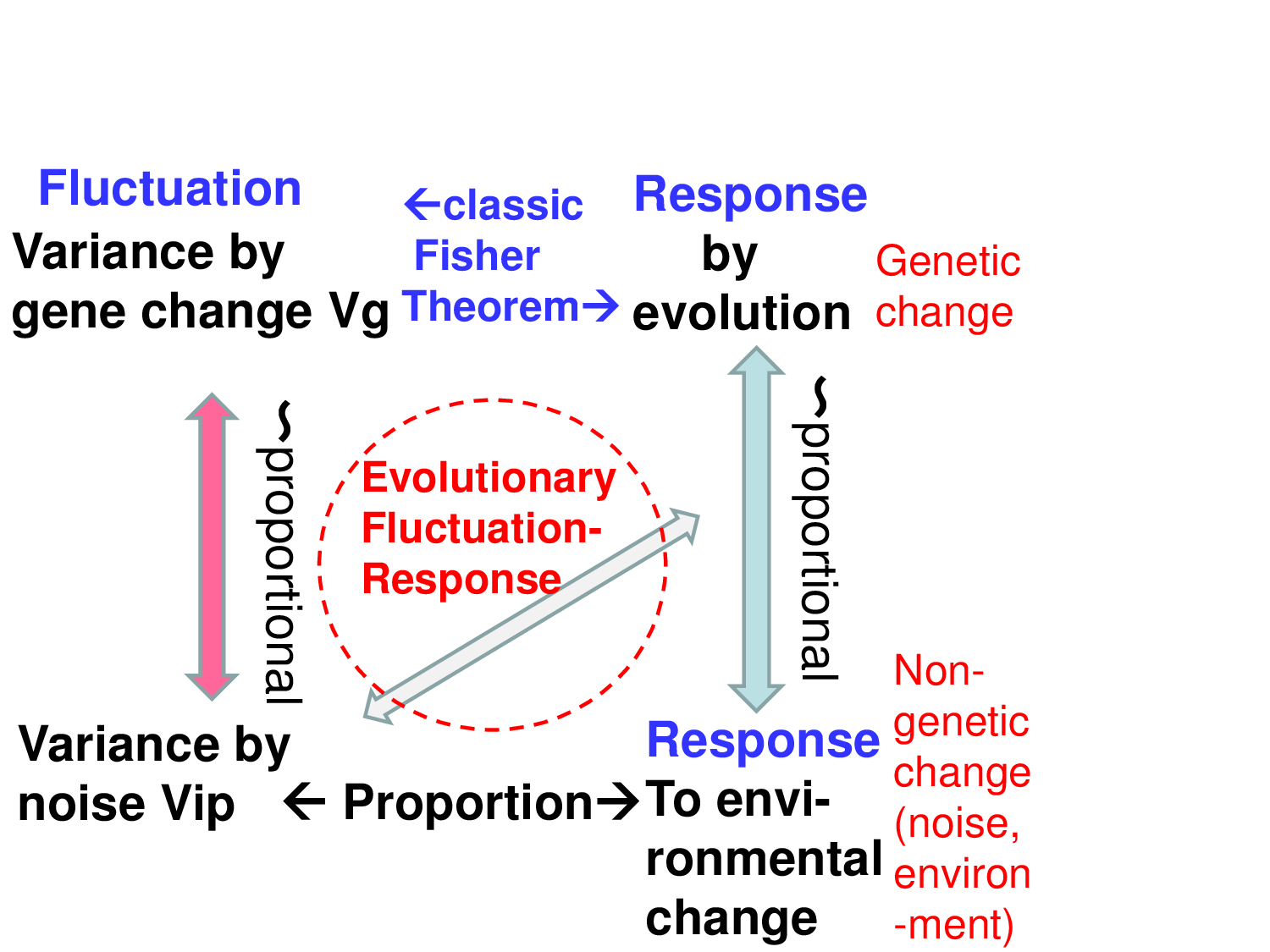}
\end{center}
\caption{
Schematic representation of  2x2 proportionality relationship between environmental response and evolutionary response, as well as between te variance $V_{ip}(i)$ and $V_g(i)$, and between fluctuation and response.
}
\end{figure}

\section{Universality}

The universality of evolutionary dimension reduction, the proportionality between responses to environmental and genetic changes, and the proportionality between genetic variance ($V_g$) and isogenic phenotypic variation ($V_{ip}$) have recently been examined in theoretical models and experiments. These are briefly reviewed here.

(I) Reaction-network and gene-regulation network models:

Dimensional reduction by evolution has been observed in several models. First, in reaction network models, phenotypic change is mainly constrained along a one-dimensional manifold by different types of fitness for selection (e.g., by selecting the concentration of a component). Second, even when environmental conditions (e.g., the concentration of external nutrients) are not fixed but fluctuate over generations, the constraint is observed within a one-dimensional manifold \cite{Sato-KK1}. 
Third, by using the gene expression dynamics model and imposing a fitness condition depending on the expression pattern of target genes, dimensional reduction and the relationship between variances are confirmed\cite{Sato-KK2023}.

(II) Evolution of protein structure: 

Through a combination of data analysis, elastic network models, and theoretical analysis, we have confirmed the relationship between $V_g$ and $V_{ip}$, as well as the dimensional reduction of protein structure. Changes in estimated structure due to conformation dynamics under noise and genetic mutations are constrained to a common low-dimensional space, typically about five dimensions, implying a strong reduction from the hundreds of residues that give rise to the high-dimensional state space \cite{Tang-KK}. A theoretical explanation is proposed there, whereas the consequence of few slow modes has been discussed earlier \cite{Togashi}, and some model studies also suggest such dimensional reduction \cite{Tlusty}.

In addition, we have also studied the evolving spin-Hamiltonian system as a simple, abstract model for the evolution of protein dynamics. 
Here, the spin variable $S_i$ takes up (+1) or down (-1) values whose dynamics is governed by the interaction with other spins, with the energy $J_{ij}S_i S_j$ between the spins $S_i$ and $S_j$. (If $J_{ij}$ is positive (negative), the two spins tend to align (oppose)).
We formulate the above interaction using a Hamiltonian ($H=-\sum_{i,j} J_{ij}S_iS_j$) for energy, with the equilibrium distribution $exp(-H/(kT))$ under temperature ($T$). By introducing the fitness determined by a particular configuration of target spins, the matrix $J_{ij}$ is evolved by mutation and selection \cite{Sakata-Hukushima-KK}. The numerical results as well as the statistical-physical calculations show that the evolutionary dimensional reduction and the $V_g$-$V_{ip}$ relation occur under an intermediate temperature range\cite{Sakata-KK,Pham-KK1}.

(III) Experimental confirmations: 

Dimensional reduction was observed in the transcriptome analysis of bacteria {\sl E. coli}, under different types of stress\cite{Mu}. Support for equation (1) was obtained by the bacterial evolution under ethanol stress, by using the transcriptome analysis of expression levels of mRNAs\cite{CFKK-Interface,Horinouchi2}. In addition, the transcriptome analysis showed that changes in expression levels of thousands of mRNAs under different antibiotics could be predicted by about seven principal components \cite{Furusawa}. The cross-fitness after evolution under different types of antibiotics is shown to be correlated with the response in prior to evolution, as predicted by the theory\cite{Sato-KK2023}. Further support for such dimensional reduction comes from  Raman analysis of cellular components \cite{Kamei}.

The proportionality between genetic and isogenic phenotypic variances across genes is also consistent with  measurement of yeast gene expressions. Landry {\sl et al.}\cite{Landry} made global measurements of isogenic expression variation and mutation-induced variance. Although they used ``mutational variance" (i.e., the rate of diffusive spread of each gene expression level upon the introduction of mutations) instead of $V_g(i)$, a high correlation between the isogenic phenotypic fluctuation ($V_{ip}(i)$) and the mutational variance was observed across all genes in yeast (see also \cite{Lehner}). More recently, the quantitative analysis of the gene expression pattern in Japanese {\sl medaka} (fish) \cite{Uchida} has suggested $V_g$-$V_{ip}$ proportionality.

The validity of the proportionality between variances is not limited to gene expressions, but can be applied to any other phenotypic trait. The correlation between the isogenic variance and the rate of evolution was confirmed in the experiment on the evolution of RNA structure\cite{Ichihashi}. Correlation between genetic and antigenic (``phenotypic") changes was noted in influenza evolution\cite{influenza}. In a series of experiments, Stearns et al.\cite{Stearns} measured the isogenic variance of five life-history traits in Drosophila melanogaster, while also measuring the genetic variance in the same five traits between different genetic lines.  Interestingly, they observed proportionality between the isogenic variance of each trait and its genetic variance. Further support for $V_g$-$V_{ip}$ proportionality can be found in the wing morphology of {\sl Drosophila} (fly) \cite{Saito,Rohner}.

(IV) Dimensional reduction in related topics: 

The dimensional reduction has been discussed in morphology of shells\cite{Raup}, beaks\cite{Brenner}, wing patterns\cite{Mani} and so forth. It has also been observed in the behavior of microorganisms\cite{JordanLeibler}, {\sl C. elegans}\cite{Stephens1,Stephens2}, development of {\sl C. elegans}\cite{Jordan} and in ecological evolution\cite{Leibler}.
One possible link to the dimensional reduction will be the {\sl sloppy parameter hypothesis}\cite{sloppy}, which suggests that many parameters used in biological models are irrelevant. 

Besides the issue in biophysics, dimension reduction is attracting much attention in learning in the brain \cite{Batista,Gallego} and neural networks \cite{Schreier},as well as in machine learning\cite{Mao}. In deep learning, the possible network configurations that can support a desired output are quite redundant, and reduction to the appropriate output space is achieved.  This is similar to our case. The genetic space (e.g., the network configuration) is high-dimensional, so there can be many ways to achieve the appropriate phenotype.  Even though the final selected states are constrained in low-dimensional space, the use of high-dimensional space is essential for both machine learning and evolution.

\section{Homeorhesis in Development and Evolution-Development Relationship}

A promising extension of dimensional reduction and $V_g$-$V_{ip}$ relationship will be in the area of evolutionary development.
To consider the robustness of the developmental process, Waddington, in the 1950s, schematically represented the developmental process by the motion of a ball along a landscape of branching valleys, as shown in Fig. 4\cite{Waddington}. The ball falls down into lower valleys that represent stable cell types, while the successive branching of the valleys represents the differentiation process. In the figure, the horizontal axis $X$ represents the cell state that gives rise to the cell type, and the depth axis $Y$ represents a slow developmental time. This picture has been revived in recent years and has received much attention.

\begin{figure}
\includegraphics[width=8cm]{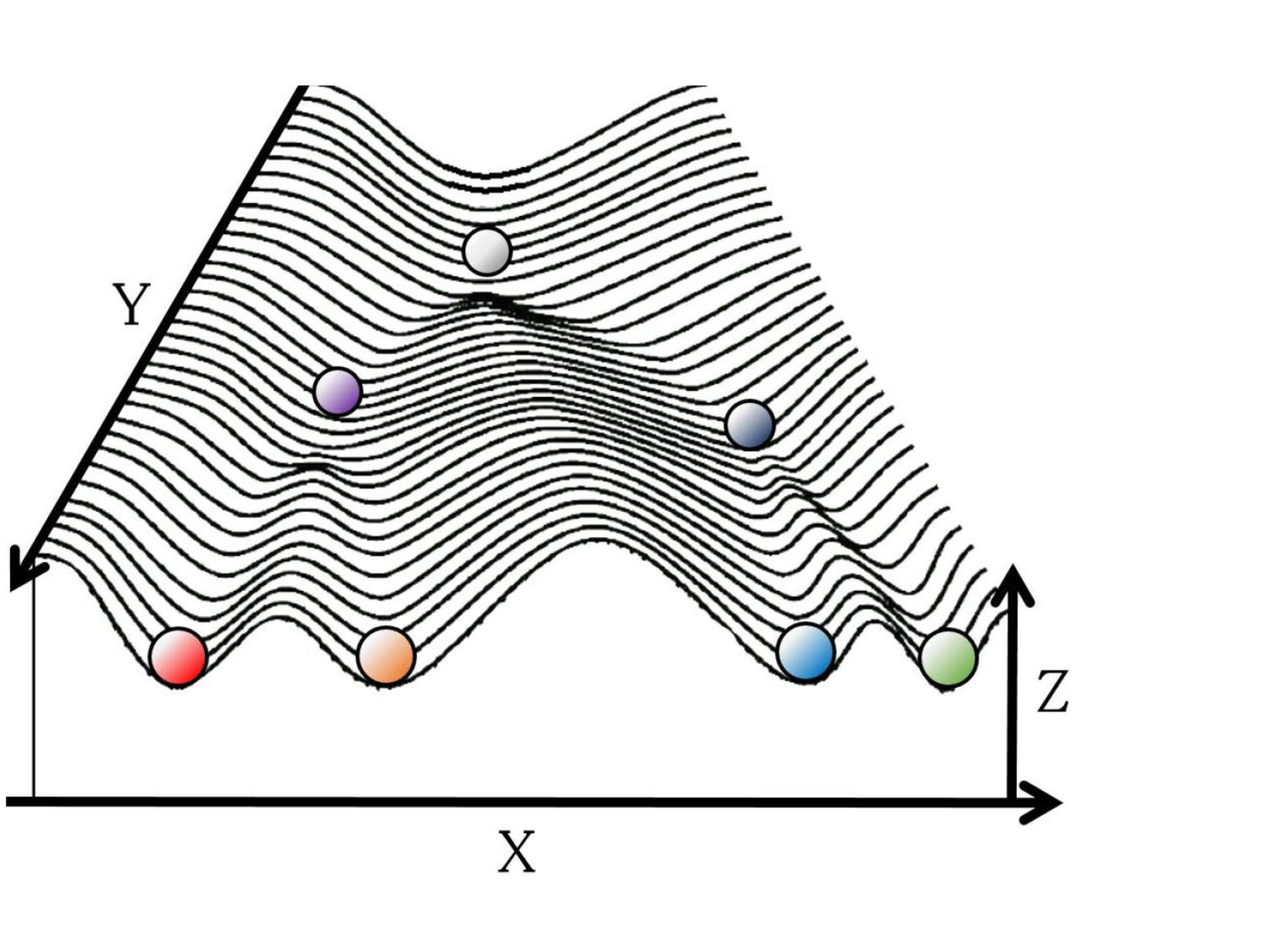}
\caption{
Schematic representation of the epigenetic landscape and cell differentiation by Waddington. Valleys in the landscapes are branched and deeper with developmental time. They correspond to differentiated cell types\cite{Waddington}.}
\end{figure}

Here, the horizontal axis represents the cellular state, possibly represented by the gene expression pattern. However, since there are thousands of genes (or components) in a cell, the cellular state involves many degrees of freedom, as already mentioned, while the horizontal axis represents only one variable. Then, how can cellular states potentially be represented by only one (or a few) variables?  This first question is again a question of dimensional reduction, which can possibly be achieved by evolution.

Since the stable state of each differentiated cell type is at the bottom of the valley, 
this may fit the attractor picture of a cellular state: 
For a given value of the depth axis $Y$, the cellular state is attracted to the bottom of the valley, where the state falls onto a fixed-point attractor, and no further change occurs. 
Here, however, the landscape itself is shaped along the depth axis $Y$ with the developmental course, with which the valleys are successively shaped and deepened in a process known as {\sl canalization}. 
Then, the second question we need to address is: what does the y-axis represent for the (slower) landscape change and what is the origin of such a slow mode\cite{Matsushita}? 

Third, developmental process is robust, i.e., how and when the valleys branch is rater robust.  Waddington coined the term {\sl homeorhesis} to discuss this robustness in the path\cite{Waddington}. This is not the stability of a state; how a developmental path is robust is the third question that remains to be answered. 

The fourth question concerns the correlation between evolution and development, which has been discussed since the time of Haeckel\cite{Haeckel}, who proposed the recapitulation theory, in which developmental processes follow a similar path to phylogenetic change. His original idea is not validated by itself, but the existence of some relationship between evolutionary and developmental changes has been noted for centuries \cite{Raff,Hall}.

In this context, the concept of the developmental hourglass has recently gained much attention, in which different species within the same phylum increase their similarity in the mid-developmental stage.  The difference in phenotypes between different species is large in the egg stage, then shrinks in the mid-developmental stage, and later increases again to have diverse adult forms. Because the bottleneck narrows in the mid-stage, it is called the hourglass hypothesis. When it was first proposed\cite{Duboule}, it was difficult to determine quantitatively, but recently it has become possible to quantitatively compare gene expression profiles at each stage. The data seem to support the hypothesis \cite{Irie,Hu,Kalinka,algae}, while further data analysis is needed to draw clear conclusions. Now, can we understand the developmental hourglass theoretically?

These four questions are now addressed in the light of the studies presented in \S 2-5, by extending the {\sl states} to {\sl paths}.

1) Dimensional reduction of developmental trajectories instead of evolutionary dimensional reduction of the state upon environmental changes.

2) Evolution of slow modes to control the developmental dynamics instead of evolution of slow modes for the intracellular process (gene expression dynamics).

3) Evolution of robustness of the developmental path rather than robustness of the final state.

4) Correlation between the time course of phenotypic variances due to noise or environmental changes and due to genetic changes, over developmental time $t$, rather than the correlation between the steady-state variances $V_g$ and $V_{ip}$.

\begin{figure*}
\begin{center}
\includegraphics[width=10cm]{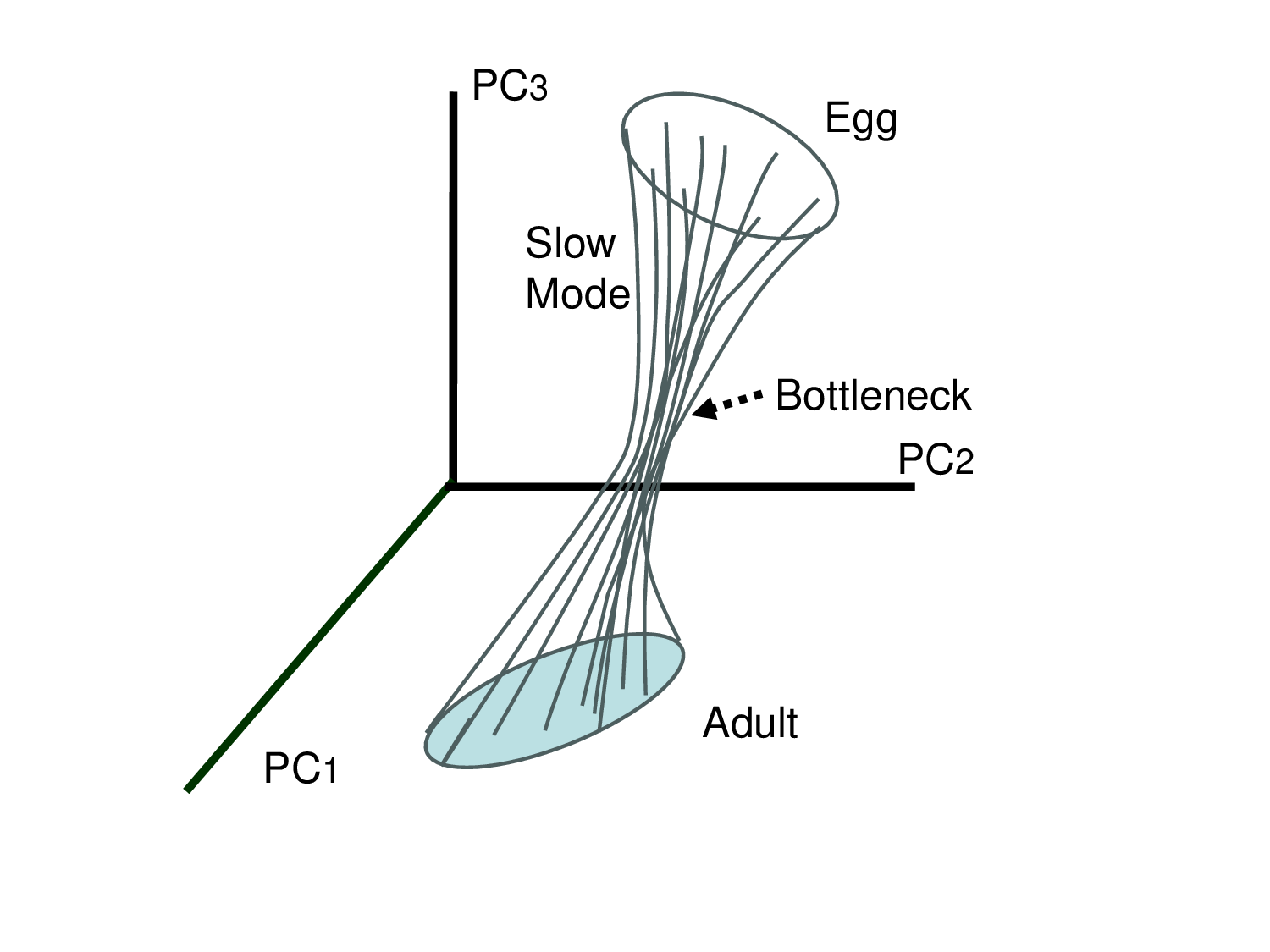}
\end{center}
\caption{
Schematic representation of developmental hourglass, robustness in developmental orbits, and dimensional reduction: The variation over initial variance or by genetic variation is decreased up to the mid-developmental stage, and then the orbits are diversified toward the adult stage. The developmental orbits are constrained in the low-dimensional space, controlled by slow modes.
}
\end{figure*}

To understand this evolutionary developmental relationship, we performed evolutionary simulations of the developmental process \cite{Kohsokabe,Kohsokabe2024} Specifically, cells follow the gene expression dynamics governed by the identical gene regulatory network. These cells are placed in one dimension, while some components are assumed to be exchanged with neighboring cells through diffusive interactions. Furthermore, external chemical components (morphogens) are supplied with a spatial gradient, providing input to the gene expression system. Under these setups, the gene expression levels (protein concentrations) of each cell reach a steady state that depends on the location of the cell, leading to pattern formation\cite{Kohsokabe}.

Let us then assume that the fitness is given by the number of final cell types defined by the expression pattern of target genes (specifically four target genes are adopted, so that $2^4$ cell types are possible according to the possibilities of all their on-off patterns.) By using the fitness and mutating the gene regulation networks over generations, the individuals with higher fitness are selected in each generation.
After a few hundred generations, the networks have evolved to reach the fittest. It turns out that during evolution, the expression of a certain gene (protein) slowly changes, controlling the expression of the output genes.  (For details, see the original paper\cite{Kohsokabe2024}). 

Now, the four questions raised above are resolved by the four points mentioned above(see Fig.5). 

A) By calculating the similarity between the spatial patterns of gene expression levels of two species that evolved from a common ancestor, a developmental hourglass was observed, i.e., the difference in gene expression patterns between the two species is minimal in the middle stage of development. This mid-stage bottleneck (or peak of similarity between the two) is prominent when the species being compared are phylogenetically close, while it becomes vague and eventually disappears as the species diverge further in the phylogeny.

B) In addition to this developmental hourglass observation, other experimentally reported features for developmental hourglass were confirmed in the model. In particular, the variances of gene expression patterns across clones are minimal at the developmental bottleneck stage: The phenotypic variances $V_{ip}(t)$ due to initial or developmental noise decrease from the egg stage to the hourglass bottleneck stage, followed by an increase that leads to diversification in later developmental processes. 

C) Behind the appearance of the developmental hourglass in the model is the acquisition of genes whose expression level changes slowly. Slow changes in the expression level of such genes control the timing of the developmental process and induce developmental robustness. They control the expression dynamics of other genes, which also facilitates evolution. In the model simulations, these slow expression dynamics have been conserved among organisms within a certain phylogenetic range of species sharing the hourglass (point 2).

D) Thus, the developmental hourglass can be related to robustness and dimensional reduction by a slow mode, as discussed in \S2: The phenotypic diversity at the egg stage, derived from the mother, has to be reduced at some point in order for the developmental process to be stable (point 3). This leads to the low-dimensional constraint on the developmental trajectory of the phenotype (point 1), which results in the reduction of the variance $V_{ip}(t)$ at mid-development, the timing of which is given by the slowly changing gene expression. By assuming that $V_{ip}(t)$ is correlated with the variance over mutants $V_g(t)$ through developmental time, the developmental bottleneck across species is expected (point 4)\cite{Kohsokabe2025}. 

\section{Collapse of dimensional reduction: {\sl Anna Karenina phenomena}?}

So far we have discussed the achievement of dimensional reduction in a biologically {\sl good} state, 
in which macro-micro consistency is satisfied, i.e., cell growth and molecular replication are balanced.
Then, it will be of interest to study if the reduction remains to be valid for cells with a bad, non-growing state which are put into  poor environmental conditions. In fact, in bacteria, 
there generally exists a transition to the dormant state, i,e., the growth-arrested state,  when the nutrient concentration is decreased\cite{dormancy,Balaban}, whereas theoretical models for the transition from an exponential growth state to a quiescent state (which cannot grow) have also been proposed\cite{Himeoka-KK,Dill}. Will the dimensional reduction collapse with this transition to the dormant state? 

In relation, a simple cell model that shows such transition to dormancy has recently been studied. 
The model consists of a catalytic reaction network with many components, where two-step elementary reaction processes with the formation of an intermediate complex $C$ are introduced\cite{YamagishiKK-2024}, such as
\begin{math}
X_i + X_j \leftrightarrow C_{ij}
\end{math}
\begin{math}
C_{ij} \rightarrow X_k+ X_j.
\end{math}

If the rate of the second reaction is much faster, the product is formed as soon as the complex is formed, so that the above reaction processes are reduced to single mass action kinetics as $X_i + X_j \rightarrow X_k+X_j$, without intermediate complex formation.  Then the model is reduced to the cell model already mentioned in \S 2-6. On the other hand, if the rate is not so fast, the intermediate complex $C$ remains for some time. These complexes can accumulate depending on the conditions, especially when the resource is limited. If such accumulation occurs, the abundances of free reactants not bound in complexes will decrease, which can hinder the reaction processes.

By numerically solving the above model with the production rate not so large, it is found that the growth rate decreases by orders of magnitude when the supplied nutrient concentration is reduced below a critical value, thus demonstrating the transition from exponential growth to the dormant phase\cite{YamagishiKK-2024}. 

This suppression of growth at the transition can be understood as follows:
In the complex reaction network with many components, there is a set of components that catalyze each other for cell growth, i.e., an autocatalytic subnetwork, while some other components attached to it are produced with the help of the autocatalytic subnetwork.  In the exponential growth phase, the abundances of the components are concentrated on the autocatalytic subnetwork, leading to the dimensional reduction. However, when the nutrient supply is reduced, the reaction flux is reduced and the complexes begin to accumulate. The catalytic reaction  then stagnates in the pathways within the autocatalytic network, increasing the fraction of flux flowing to other components.

In fact, an increase in the diversity of chemical components is observed as the cell enters a non-growing quiescent state. This increase in diversity suggests a violation of dimensional reduction in an exponential growth state.
In a bad state, the dimensional reduction achieved for a robust state with consistency between macro and micro levels may collapse, leading to diversity in components. Such collapse of dimensional reduction may be also seen in the microorganism behavior\cite{JordanLeibler}, or possibly in cancer cells with possible increase in the phenotypic variances\cite{cancer}. Probably, the first sentence of Tolstoy's {\sl Anna Karenina}, ``Happy families are all alike; every unhappy family is unhappy in its own way'', can be applied to cells or biological systems.

\section{Summary}

In this paper, we have reviewed recent advances in evolutionary dimension reduction. Phenotypes in biological systems are generally robust to perturbations whereas they maintain plasticity, so that they are changed by evolution. The phenotypic changes are originally be high-dimensional, but to satisfy this robustness and plasticity, they are constrained to a low-dimensional manifold (space). In this case, the responses of phenotypes to environmental changes, to stochastic perturbations, and to genetic changes follow dominantly along this low-dimensional manifold, and thus the responses to environmental and genetic changes, as well as fluctuations to noise, and to genetic mutations are strongly correlated, allowing the prediction of phenotypic evolution. The extension of this dimensional reduction theory to the development-evolution relationship is briefly reviewed. Finally, a possible collapse of the reduction in a dormant (non-growing) cellular state is briefly discussed.

This research was partially supported by and the Novo Nordisk Foundation (NNF21OC0065542) and Grant-in-Aid for Scientific Research (A) (20H00123) from the Ministry of Education, Culture, Sports, Science and Technology (MEXT) of Japan. The author would like to thank Chikara Furusawa, Takahiro Kohsokabe, Takuya Sato, Ayaka Sakata, Qian-Yuan Tang, and Tetsuhiro S. Hatakeyama for useful discussions.

\end{document}